\documentclass[prl,aps,twocolumn,superscriptaddress]{revtex4}

\usepackage{graphicx,epic,eepic,epsfig,amsmath,latexsym,amssymb,verbatim,subfigure,color}

\begin{document}
\newcommand{\ket}[1]{| #1 \rangle}
\newcommand{\bra}[1]{\langle #1 |}

\title{A simple family of nonadditive quantum codes}
\author{John A. Smolin}
\email{smolin@watson.ibm.com}
\affiliation{IBM T.J. Watson Research Center, Yorktown Heights, NY 10598, 
U.S.A.}
\author{Graeme Smith}
\email{gsbsmith@gmail.com}
\affiliation{Department of Computer Science, University of Bristol, Bristol, BS8 1UB, UK}

\author{Stephanie Wehner}
\email{wehner@cwi.nl}
\affiliation{C.W.I., Kruislaan 413, 1098 SJ Amsterdam, The Netherlands}

\begin{abstract}
Most known quantum codes are additive, meaning the codespace can be
described as the simultaneous eigenspace of an abelian subgroup of the
Pauli group.  While in some scenarios such codes are strictly
suboptimal, very little is understood about how to construct
nonadditive codes with good performance.  Here we present a family of
nonadditive quantum codes for all odd blocklengths $n$, that has a
particularly simple form.  Our codes correct single qubit erasures
while encoding a higher dimensional space than is possible with an
additive code or, for $n\geq 11$, any previous codes.
\end{abstract}
\maketitle

Quantum error correcting codes will be essential if large-scale
quantum computers are ever to be built.  Nearly all known quantum
codes are stabilizer (or additive) codes, in the framework of
\cite{gottesman}, but it has been known for some time that nonadditive
codes can perform better \cite{RHSS}.  So far this has only been shown
for distance 2 quantum codes.  A distance 2 code corrects
any single qubit erasure, or alternatively, can be used to detect a
single qubit error with unknown location.  Particularly in this second
capacity, such codes may be quite important
in the ancilla preparation phase of fault-tolerant quantum 
computing \cite{knill}.
Furthermore, distance 2 codes promise to shed light on the structure of
general quantum codes due to their simplicity.

It is shown in \cite{gf4} that the best additive distance-2 codes are
$[n,n-2,2]$ for $n$ even and $[n,n-3,2]$ for $n$ odd, where we have used the notation $[n,k,d]$ to indicate an $n$ qubit additive code with distance $d$ and $k$ encoded qubits. In
\cite{RainsDistance2} it was shown that that these codes are optimal
for even $n$.  For $n=5$, a $((5,6,2))$ nonadditive code was found in
\cite{RHSS}, where we have used the notation $((n,K,d))$ to indicate a
distance $d$ code of size $n$, protecting a $K$ dimensional code space, in 
analogy with the standard classical notation of $(n,K,d)$ denoting an $n$
bit code with $K$ codewords and distance $d$.  This code, together with the family of $((n,3\cdot
2^{n-4},2))$ codes it generates \cite{RainsDistance2}, gives the only
performance improvement from nonadditive codes known for qubits. 
Here, we present a family of new nonadditive codes that improve on all known constructions.
Our codes correct single qubit erasures while encoding a higher
dimensional space than is possible with any additive code and, for $n \geq 11$, 
any nonadditive code known to date. In particular, we show how to construct $((4k + 2l + 3, M_{k,l},2))$-codes
where $M_{k,l} \approx 2^{n-2}(1 - \sqrt{2/(\pi(n-1))}$. Our construction is not restricted to qubits, but
can be extended to higher dimensional systems.

It is shown in \cite{RainsDistance2} that for odd $n$ the largest
codespace dimension $K_{\rm max}$ is bounded by
\begin{equation}
K_{\rm max} \le 2^{n-2}\left(1-\frac{1}{n-1}\right)\ .
\label{kmax}
\end{equation}
While we would like to approach this bound, at least for large $n$, we
achieve a more modest goal. Our code is asymptotically almost optimal, but
the rate of convergence is weaker due to the square root in $M_{k,l}$.


\vspace{-.1in}
\section{The code for $n=5$} 
We first describe our construction for
$n=5$ and then show how to extend it to general odd $n$.
We obtain our quantum codes from specific classical codes. 
Consider the following five five-bit strings:
\begin{equation}
\nonumber \mathbf{x}^{(j)}\ \ 0\le j<5
\end{equation}
whose $i$th bits are given by
\begin{equation}\nonumber
x_i^{(j)}=\delta_{ij}\ \ 0\le i<5\ .
\end{equation}
These, together with their bitwise complements
$\bar{\mathbf{x}}^{(j)}$, form a classical $(5,10,2)$ code.  This can
be seen easily as every codeword is either weight 1 or weight 4, and
single bit errors necessarily change the weight of a codeword by $1$,
taking it out of the codespace. In the following, we will call a code
\emph{self-complementary} if the complement of each codeword is also 
in the code.

A quantum code must detect more than simply bitflips.  It must detect
the errors $X={{0\ 1}\choose {1\ 0}}$ (``amplitude'' errors, analogous
to the classical bitflip), $Z={{1\ \ \,0}\choose {\,0\ -1}}$ (errors in
phase) and $Y=XZ$ (when both errors occur)\footnote{We omit factors of
$\sqrt{-1}$ in the Pauli matrices.
}.  
It will be sufficient to show for our code that any of these
errors map code vectors to some vector orthogonal to the codespace.
The full necessary and sufficient error-correction conditions were
derived in \cite{BDSW,KLF}. 

We now show how to turn the classical code above into a quantum code.
The five basis vectors of our quantum code are related to the
classical code by taking superpositions of codewords with their
complements:
\begin{equation}\nonumber
v^{(j)}=\ket{\mathbf{x}^{(j)}}+\ket{\bar{\mathbf{x}}^{(j)}}\ \ 0\le j<5
\end{equation}
\begin{eqnarray}
\nonumber v^{(0)}&=&\ket{10000}+\ket{01111}\\
\nonumber v^{(1)}&=&\ket{01000}+\ket{10111} \\
\nonumber v^{(2)}&=&\ket{00100}+\ket{11011}\\
\nonumber v^{(3)}&=&\ket{00010}+\ket{11101}\\
\nonumber v^{(4)}&=& \ket{00001}+\ket{11110}.
\end{eqnarray}
What is the effect of single qubit errors?
An $X$ error on the $i$th qubit acts on the codewords as
\begin{equation}
X_i v^{(j)} =\ket{\mathbf{x}^{(i)}\oplus \mathbf{x}^{(j)}} + 
\ket{{\mathbf{x}}^{(i)}\oplus\bar{\mathbf{x}}^{(j)} },
\label{xes}
\end{equation}
so that for $i\ne j$ the $X_i v^{(j)}$'s are all the ``+'' superpositions of
weight two vectors with their weight three complements, while for
$i=j$ we always have the state $\ket{00000}+\ket{11111}$, the 
superposition of the weight zero vector with its weight five complement.
These are all orthogonal to the codespace 
since they are constructed entirely of superpositions of weights not in
the codewords, which is a direct result of having started with a classical
distance $d=2$ code.

Turning our attention to the $Z$ errors, we find
\begin{equation}\nonumber
Z_i v^{(j)} = (-1)^{\delta_{ij}}(\ket{\mathbf{x}^{(j)}}-
\ket{\bar{\mathbf{x}}^{(j)}}).
\end{equation}
These five vectors are the original code vectors but with a ``-''
as relative phase and sometimes an overall phase.  They
are all orthogonal to the code vectors and to one another, as well as
to the $X$ error states of Eq.(\ref{xes}).

Finally, $Y$ errors act on the codewords according to 
\begin{equation}\nonumber
Y_i v^{(j)} = (-1)^{\delta_{ij}}(\ket{\mathbf{x}^{(i)}\oplus \mathbf{x}^{(j)}} 
- \ket{{\mathbf{x}}^{(i)}\oplus\bar{\mathbf{x}}^{(j)} } ).
\end{equation}
These are again orthogonal to the code space as they consist of 
superpositions of weights not included in the original code, and also to the image of the code under
$X$ and $Z$ errors (the $X$'s due to the relative phase, and the $Z$'s due to being
made of the wrong weight strings).

We emphasize that this code is not a subcode of the $((5,6,2))$ code
in \cite{RHSS}.  Indeed, it is not a subcode of {\em any} $((5,6,2))$
code, since there are no vectors orthogonal to the codespace with
distance 2 to all five vectors.  To see this, observe that the set of
vectors obtained by single qubit errors on the code vectors spans the
entire space orthogonal to the codespace.  Since every vector outside
the codespace might be mistaken for a single error on some state in
the code, no such vector can be added to our code while maintaining a
minimum distance of 2.

\section{The code for odd $n$}
We now show how to extend our construction to any odd $n$.
The method we have used is quite general, and is summarized in the
following lemma, whose proof follows immediately from the reasoning
above:

\noindent {\bf Lemma 1:} Any self-complementary $(n,K,d>1)$ classical code 
leads to a $((n,K/2,2))$ quantum code by pairing up codewords with 
their complements in superposition.

Our construction again starts from a classical code, which is obtained as follows:
Choose an ordering of the weight $i$ bitstrings of length
$n$ and let $\mathbf{w}^{(i,j,n)}$ be the $j$th such string, where $0
\leq j < {n\choose i}$.  Letting $n = 4k+2l+3$, we consider the
classical distance-2 codes indexed by $(k,l)$ whose codewords (indexed
by $(i,j)$) are
\begin{eqnarray}\label{vs}
\mathbf{v}^{(i,j)}_{(k,l)}=\mathbf{w}^{(2i+l,j,4k+2l+3)}&& l=0,1\\
\nonumber &&0\le i \le k \\
\nonumber &&0\le j<{4k+2l+3 \choose 2i+l}
\end{eqnarray}
together with their complements $\bar{\mathbf{v}}^{(i,j)}_{(k,l)}$.  
Note that our $(5,10,2)$ code
is the special case of (\ref{vs}) with $k=0,l=1$. 

Using Lemma 1, we now turn this code into a 
$((n=4k+2l+3,M_{(k,l)},2))$ quantum code, spanned by
\begin{eqnarray}\nonumber
\psi^{(i,j)}_{(k,l)}=\ket{\mathbf{v}^{(i,j)}_{(k,l)}}+
\ket{\bar{\mathbf{v}}^{(i,j)}_{(k,l)}}
&& l=0,1\\
\nonumber &&0\le i \le k \\
\nonumber &&0\le j<{4k+2l+3 \choose 2i+l}.
\end{eqnarray}
Let $C_n$ denote our code.

It remains for us to
count $M_{(k,l)}$, the total number of code vectors.  We have
for $l \in \{0,1\}$ that
\begin{equation}\nonumber
M_{(k,l)}=\sum_{i=0}^k {4k+2l +3 \choose 2i + l}=2^{4k+2l+1}-\frac{1}{2}{4k+2l+2 
\choose 2k+l+1},
\end{equation}
where we have evaluated the sum using Pascal's first identity 
${n \choose m}={n-1 \choose m} + {n-1 \choose m-1}$ \cite{pascal}.

As mentioned above, $((n,3\cdot 2^{n-4},2))$ codes were constructed in
\cite{RainsDistance2}.  For $n \leq 9$, these codes encode more
elements than ours, while for $n\geq 11$ our codes have a larger
codespace.  Evaluating $M_{(k,l)}$ as $n\rightarrow \infty$ gives
\begin{equation}\nonumber
M_{(k,l)}=2^{n-2} \left( 1-\frac{{n-1 \choose \frac{n-1}{2}}}{2^{n-1}}\right)
\approx 2^{n-2}\left( 1- \sqrt{\frac{2}{\pi (n-1)}}\right)
\end{equation}
allowing us to asymptotically encode $n-2- \frac{1}{\ln
2}\sqrt{\frac{2}{\pi(n-1)}}$ qubits.  While the rate of convergence
to may be suboptimal ($O(\frac{1}{\sqrt{n}})$ vs. $O(\frac{1}{n})$),
in light of Eq.~(\ref{kmax}) the resulting limit of $n-2$ encoded
qubits cannot be surpassed and our code is asymptotically close to optimal.\\

\vspace{-.1 in}
\section{The automorphism group}

The automorphism group of a code is the group of unitaries that map
the codespace to itself and consist of the composition of local
unitaries on each qubit and a permutation of qubits.  The
automorphisms of a code characterize the code's symmetries. Since
they correspond to the logical operations that can be applied
transversally, they are relevant for fault tolerant quantum computation.
The following lemma gives the automorphism group of our code.\\

\noindent{\bf Lemma 2:}
The automorphisms of $C_n$ are exactly
\begin{equation}
\left(X^{\otimes n}\right)^{b} Z^{\mathbf{f}} \circ \pi_n \equiv \left(X^{\otimes n}\right)^{b} \left(\bigotimes_{l=1}^n Z^{f_l}\right) \circ \pi_n,
\end{equation}
with $|\mathbf{f}|$ even, $b \in \{0,1\}$ and $\pi_n \in S_n$ is a
permutation.\\

This is most easily proved as a consequence of the following characterization of the
projector onto $C_n$.\\

\noindent{\bf Lemma 3:}  The projector onto $C_n$ is given by
\begin{equation}
\label{Eq:projector}
P_{C_n} = \frac{1}{2^n} \left( I^{\otimes n} + X^{\otimes n}\right) 
\sum_{s=0}^{\frac{n-1}{2}} K^{(2s)} \left(\sum_{|\mathbf{x}| = 2s} 
Z^{\mathbf{x}}\right)
\end{equation}
where we have let $K^{(2s)}= \sum_{i=0}^{k}K^{(2s)}_{2i{+}l}$, as well as 
$K^{(2s)}_{2i{+}l}=2 \sum_{t=0}^{2i+l}\binom{2s}{t}\binom{n-2s}{2i{+}l{-}t}(s{-}t)$
for $ s>0$ and $K^{(0)}_{2i{+}l} = \binom{n}{2i{+}l}$.

\noindent{\em Proof:}
Letting $P_{2i+l} = 
\sum_{j} \ket{\psi^{(i,j)}_{(k,l)}}\!\bra{\psi^{(i,j)}_{(k,l)}}$, 
we have
\begin{equation}
P_{2i+l} = \frac{1}{2}\sum_{|\mathbf{w}| = 2i+l}\sum_{b_1b_2=0}^1  (X^{\otimes n})^{b_1}\ket{\mathbf{w}}\!\bra{\mathbf{w}}(X^{\otimes n})^{b_2},
\end{equation}
which, using the fact that $\ket{\mathbf{w}}\!\bra{\mathbf{w}} = \bigotimes_{l=1}^n \left(\frac{I+(-1)^{w_l}Z}{2} \right)$, is equal to  
\begin{equation}\nonumber
\frac{1}{2^n}\sum_{\mathbf{|x|}{\ \rm even}} 
\left( \sum_{|\mathbf{w}| = 2i+l} (-1)^{\mathbf{x} \cdot \mathbf{w} }\right)
\left( I^{\otimes n} + X^{\otimes n}\right)Z^{\mathbf{x}}
\end{equation}
\begin{equation}
 = \frac{1}{2^n}\sum_{s=0}^{\frac{n-1}{2}}K^{(2s)}_{2i+l}\left( I^{\otimes n} + X^{\otimes n}\right)
 \left(\sum_{|\mathbf{x}| = 2s} Z^{\mathbf{x}}\right).
\end{equation}
Summing over $i=0\dots k$ proves the claim. $\Box$


\noindent{\em Proof (of Lemma 2):} Since our code is permutation
invariant, we only need to show that the unitaries of the form
$\bigotimes_{l=1}^n U_l$ leaving $C_n$ invariant are exactly of the
form $Z^{\mathbf{f}}$ or $X^{\otimes n}Z^{\mathbf{f}}$ with
$|\mathbf{f}|$ even.  To see this, note that Eq.~(\ref{Eq:projector})
has terms of two types: those of weight (the number of non-identity Paulis) less than $n$, namely

\begin{equation}
\frac{1}{2^n} \sum_{s=0}^{\frac{n-1}{2}} K^{(2s)} 
\left(\sum_{|\mathbf{x}| = 2s} Z^{\mathbf{x}}\right),
\end{equation}
and those of weight $n$.  Since conjugation by local unitaries does
not change the weight of a Pauli operator (and indeed, acts trivially
on identity factors), if we wish $P_{C_n}$ to be invariant under
conjugation by $\bigotimes_{l=1}^n U_l$, we must have
\begin{equation}
\left(\bigotimes_{l=1}^n U_l\right) Z^{\mathbf{x}} 
\left(\bigotimes_{l=1}^n U_l^\dagger\right) =  Z^{\mathbf{x}} 
\end{equation}
whenever $|\mathbf{x}|$ is even.  In other words, $\bigotimes_{l=1}^n
U_l$ must commute with $Z^{\mathbf{x}}$ for all even weight
$\mathbf{x}$.  This implies $\bigotimes_{l=1}^n U_l = (X^{\otimes
n})^bZ^{\mathbf{f}}$ for $b \in \{0,1\}$.  To see that we must have
$|\mathbf{f}|$ even, note that $ (X^{\otimes n})^bZ^{\mathbf{f}}$
commutes with $ \frac{1}{2^n} \sum_{s=0}^{\frac{n-1}{2}} K^{(2s)}
\left(\sum_{|\mathbf{x}| = 2s} Z^{\mathbf{x}}\right)$, so that to
commute with $P_{C_ n}$ it must also commute with $X^{\otimes
n}$. $\Box$

\vspace{-.05 in}
\section{Conclusion}
Our family of codes can be extended to Heisenberg-Weyl type errors
on higher dimensions than qubits with exactly the same counting
as before.  Instead of using classical codewords paired with their
complements, one uses codewords of the form
\begin{equation}
\ket{\mathbf{v}}+X^{\otimes n} \ket{\mathbf{v}} + 
(X^{\otimes n})^2 \ket{\mathbf{v}} \ldots 
(X^{\otimes n})^{D-1} \ket{\mathbf{v}}
\end{equation}
in $D$ dimensions, where $X\ket{i}=\ket{(i+1) \mod D}$ defines the
$D$-dimensional $X$ operator.  The same classical codewords $\mathbf{v}$ of Eq.~(\ref{vs})
generate quantum $((4k+2l+3,M_{(k,l)},2))$-codes with $M_{(k,l)}$
as before.

So far we have not been able to find similar constructions for
higher distances as Lemma 1 is specific to distance 2, but we
are hopeful our work will inspire new thinking on nonadditive codes.

\noindent{\em Acknowledgments:} Thanks to A.W. Cross and S. Bravyi for
helpful discussions. JAS thanks ARO contract DAAD19-01-C-0056 and GS
thanks the UK Engineering and Physical Sciences Research Council.
SW thanks IBM Watson for their hospitality, and the NWO vici grant 2004-2009
and EU project QAP (IST-2005-15848) for support.

\end{document}